\documentclass[12pt,preprint]{aastex}
\begin{document}
\title{Comparison between the Blue and the Red Galaxy Alignments detected 
in the Sloan Digital Sky Survey}
\author{Jounghun Lee}
\affil{Department of Physics and Astronomy, FPRD, Seoul National University, 
Seoul 151-747, Korea} 
\email{jounghun@astro.snu.ac.kr}
\author{Ue-Li Pen}
\affil{Canadian Institute for Theoretical Astrophysics, Toronto, ON M5S 3H8, 
Canada} 
\email{pen@cita.utoronto.ca}

\begin{abstract}
We measure the intrinsic alignments of the blue and the red galaxies 
separately by analyzing the spectroscopic data of the Sloan Digital Sky Survey 
Data Release 6 (SDSS DR6). For both samples of the red and the blue galaxies 
with axial ratios of $b/a\le 0.8$, we detect a $3\sigma$ signal of the 
ellipticity  correlation in the redshift range of $0\le z\le 0.4$ for $r$-band 
absolute (model) magnitude cut of $M_{r}\le -19.2$ (no $K$ correction). 
We note a difference in the strength and the distance scale for the red and 
the blue galaxy correlation $\eta_{2D}(r)$: For the bright blue galaxies, 
it behaves as a quadratic scaling of the linear density correlation of 
$\xi(r)$ as $\eta_{2D}(r)\propto\xi^{2}(r)$ with strong signal detected only 
at small distance bin of $r\le 3h^{-1}$Mpc. 
While for the bright red galaxies it follows a linear scaling as 
$\eta_{2D}(r)\propto\xi(r)$ with signals detected at larger distance out to 
$r\sim 6h^{-1}$Mpc. We also test whether the detected correlation signal is 
intrinsic or spurious by quantifying the systematic error and find that the 
effect of the systematic error on the ellipticity correlation is negligible. 
It is finally concluded that our results will be useful for the weak lensing 
measurements as well as the understanding of the large scale structure 
formation.
\end{abstract}
\keywords{cosmology:observations --- methods:statistical --- 
large-scale structure of universe}

It has been long suspected that the galaxies may have preferential directions 
in the orientations of their elliptical shapes (or spin axes) with being 
locally correlated between the neighbors, which is often called the galaxy 
intrinsic alignments. Since the galaxy intrinsic alignments could in principle 
contaminate weak lensing signals 
\citep[e.g.,][and references therein]{hir-etal07}, the measurement of its 
existence and strength from observations has been a prior interest especially 
of the weak lensing community.  
  
It has been claimed by \citet{man-etal06} that no signal of ellipticity 
correlation between the neighbor galaxies was detected in the spectroscopic 
data from the Sloan Digital Sky Survey \citep[SDSS,][]{yor-etal00}, although 
they mentioned that their results are statistically consistent with 
previous observational evidences for the existence of the galaxy intrinsic 
alignments \citep{bro-etal02,hey-etal04}. 
Extrapolating their results to redshift $z=1$ relevant for the weak lensing 
shear, they have drawn a tentative conclusion that the intrinsic ellipticity 
correlations between the neighbor galaxies would not cause any significant 
errors for the estimation of the power spectrum amplitude of the initial 
density field.

Very recently, however, a numerical analysis based on the high-resolution 
N-body simulation has demonstrated with high statistical power that strong 
ellipticity correlations between the dark matter halos exist at large 
distances out to $10h^{-1}$Mpc \citep{lee-etal07}. Therefore, it is necessary 
and important to reexamine observationally the existence and significance 
of the galaxy intrinsic alignments by analyzing the SDSS data.

The physical mechanism for the generation of the galaxy intrinsic alignments 
differ between the blue and the red galaxies. For the blue spiral galaxies 
which are usually rotationally supported, the intrinsic alignments refer to 
the correlations of the spin axes originated from the initial tidal field. 
The linear tidal torque theory explains that the intrinsic correlation 
should exist only locally between close galaxy pairs, whose functional 
form is predicted to be a quadratic scaling of the linear density 
correlation function \citep{pen-etal00}. 

While for the bright red galaxies which are not flattened by rotation, 
the intrinsic alignments refer to the correlations between the major principal 
axes of the galaxy's elliptical shapes which are determined in observations 
in terms of the anisotropy in their stellar distribution. 
It has yet to be fully understood what the origin of the anisotropic steller 
distribution of the red galaxies is and how it is related to the shapes of 
the host halos. But the standard model based on the cosmic web theory 
suggests that it may be due to the anisotropic accretion and infall of 
materials and gas into the host halo along the halo's major principal axes 
which are in turn elongated with the large scale filamentary matter 
distribution \citep{wes94,bon-etal96}. Given this anisotropic infall scenario, 
the red galaxies may have built-in memory of the large-scale filamentary 
structures and thus their ellipticities might be correlated on larger scales 
than the spin axes of the blue galaxies. 

Even for the blue galaxies, it was claimed that the growth of the 
non-Gaussianity in the density field would cause large-scale correlations 
of their spin axes \citep{hui-zha02}.  To account for the large-scale 
intrinsic correlations between the neighbor galaxies, we use the following 
formula: 
\begin{equation}
\label{eqn:3d}
\eta_{3D}(r) \approx \frac{1}{6}a^2_{\rm l}\frac{\xi^{2}(r;R)}{\xi^{2}(0;R)} 
+ \varepsilon_{\rm nl}\frac{\xi(r;R)}{\xi(0;R)},
\end{equation}
where $r$ is the three dimensional separation distance of a galaxy pair 
and $\xi(r;R)$ is the two-point correlation function of the linear density 
field smoothed on the Lagrangian galactic scale $R$. 
Here, the two correlation parameters $a_{\rm l}$ and $\varepsilon_{\rm nl}$ 
represent the strength of the small-scale and the large-scale correlation, 
respectively. If $\varepsilon_{\rm nl}=0$ and $a_{\rm l}> 0$,  the galaxy 
intrinsic correlation will behave as a quadratic scaling of the density 
correlation function so that $\eta_{3D}(r)$ should rapidly diminishes 
to zero as $r$ increases, as the linear tidal torque theory predicts. 
In contrast, if $\varepsilon_{\rm nl} > 0$ and $a_{\rm l}=0$, the galaxy's 
intrinsic correlation will follow a linear scaling of $\xi(r)$ so that 
$\eta_{3D}(r)$ should be non-negligible even at large distance.

In fact, we have suggested this formula (eq.[\ref{eqn:3d}]) in our previous 
work \citep{lee-pen07} to account for the non-Gaussianity effect on the 
galaxy's spin-spin correlations which are tidally induced. 
Here we use the same analytic model to describe the intrinsic 
ellipticity-ellipticity correlations of the galaxies, assuming that since 
the anisotropic infall of materials that resulted in the ellipticities of 
the galaxies are also nonlinear manifestation of the tidal field and thus 
can be approached with the same analytic techniques.
Throughout this Letter, the galaxy intrinsic alignments refer to the 
intrinsic ellipticity-ellipticity alignments of the galaxies, unless 
otherwise stated. 
 
In practice, what can be measured is the intrinsic alignments not of the 
galaxy's three dimensional shapes but of its two dimensional shapes projected 
onto the plane of the sky. According to \citet{lee-pen01},  if the two 
dimensional projection effect is taken in to account properly, the 
correlation parameter has to be multiplied by a factor of $5/4$. Thus, the 
two dimensional projected intrinsic correlation function is modified 
from equation (\ref{eqn:3d}) simply as 
\begin{equation}
\label{eqn:2d}
\eta_{2D}(r) \approx 
\frac{25}{96}a^2_{\rm l}\frac{\xi^{2}(r;R)}{\xi^{2}(0;R)} + 
\frac{5}{4}\varepsilon_{\rm nl}\frac{\xi(r;R)}{\xi(0;R)}.
\end{equation}
Note here that although $\eta_{2D}(r)$ represents the intrinsic correlations  
of the galaxy's two dimensional projected ellipticities, the distance $r$ 
represents the three dimensional separation. It is also worth mentioning 
here that for the blue galaxies the intrinsic correlations of the galaxy's 
two dimensional projected ellipticities are in fact same as that of the 
galaxy's two dimensional projected spin axes since the spin axes are 
believed to be orthogonal to the galaxy's major axes.

To measure $\eta_{2D}(r)$ from the real universe, we use a dataset of galaxies 
at redshift $0\le z \le 0.4$ downloaded from the SDSS DR6 website 
(http://www.sdss.org/dr6/). A total of $640647$ galaxies are found to be in 
this redshift range. For each galaxy, we obtain information on right ascension 
($\alpha$), declination ($\delta$), redshift ($z$), position angle ($p$), 
major-to-minor axis ratios ($b/a$), the $u-g$ and $g-r$ colors, and $r$-band 
model magnitude ($M_{r}$). Among $640647$ galaxies, the information on $p$ 
and $b/a$ are available only for $639727$ galaxies, from which we also select 
only those galaxies with $b/a\le 0.8$ since for the nearly face-on galaxies 
with $b/a > 0.8$ the measurements of the direction of their major axes may 
suffer from large uncertainty. A total of $434849$ galaxies are found to 
satisfy the criterion of $b/a\le 0.8$.

Using the empirical $u-r$ color separator suggested by 
\citet{str-etal01} for the SDSS data, we classify the galaxies 
into the red ($u-r>2.22$) and the blue ($u-r\le 2.22$) sample. 
Then, we select only bright galaxies by applying the absolute magnitude 
cut of $M_{r}\le -19.2$. 
Since the SDSS calibration uses the inverse hyperbolic sine magnitudes 
(asinh) \citep{lup-etal99}, we calculate $M_{r}$ as the asinh magnitudes 
using the information on the softening parameter given in the SDSS web site.
Finally, we end up having two SDSS samples which consist of a total of 
$87188$ blue and $283972$ red galaxies brighter than this magnitude cut 
of $M_{r}\le -19.2$. 

For each galaxy, we determine the direction of the major axis, $\hat{\bf e}$, 
in the equatorial coordinate system from the given information on $p$, 
$\alpha$ and $\delta$: Let us consider two galaxies in a given sample 
whose major axis directions are found as 
$\hat{\bf e}_{i}$ and $\hat{\bf e}_{j}$. First, we  calculate 
the unit projected separation vector as 
$\hat{\bf d}\equiv {\bf d}/\vert{\bf d}\vert$ with 
$\hat{\bf d}\equiv \hat{\bf r}_{i}-\hat{\bf r}_{j}$, where 
$\hat{\bf r}_{i}$ and $\hat{\bf r}_{j}$ represent the unit position 
vectors of the $i$-th and the $j$-th galaxy, respectively. The separation 
vectors projected onto the plane of the sky at the position of the two 
galaxies are given as $\hat{\bf d}_{i}\equiv \hat{\bf d} - 
(\hat{\bf d}\cdot\hat{\bf r}_{i})\hat{\bf r}_{i}$ and $\hat{\bf d}_{j}\equiv 
\hat{\bf d} - (\hat{\bf d}\cdot\hat{\bf r}_{j})\hat{\bf r}_{j}$, 
respectively. Then, the intrinsic correlations of the two dimensional 
projected ellipticities between the galaxies with separation of $r$ 
is calculated as	
\begin{equation}
\label{eqn:gam}
\eta_{2D}(r) \equiv \sum_{i,j}
\cos^{2}\left(\gamma_{i}-\gamma_{j}\right) - \frac{1}{2}, 
\qquad {\rm with} \qquad
\gamma_{i}\equiv \tan^{-1}\left[ \frac{(\hat{\bf r}_{i}\times 
\hat{\bf d}_{i})\cdot\hat{\bf e}_{i}}{\hat{\bf d}_{i}\cdot\hat{\bf e}_{i}}
\right].
\end{equation}
Here, the angle $\gamma_{i}$ represents the direction of $\hat{\bf e}_{i}$ 
projected onto the plane of sky at the position of ${\bf r}_{i}$. Note that 
for a galaxy pair with small distance, the quantity
$\Delta\gamma\equiv \gamma_{i}-\gamma_{j}$ is actually identical to the 
difference in the position angle $\Delta p$. To determine the three 
dimensional separation distance, $r$, of a galaxy pair,  we assume a flat 
$\Lambda$-dominated cosmology with matter density $\Omega_{m}=0.25$, 
vacuum energy density $\Omega_{\Lambda}=0.75$ and Hubble constant 
$H_{0}=100 h$km s$^{-1}$Mpc$^{-1}$ with $h=1$ \citep{bla-etal03}. 

Then, we measure $\eta_{2D}(r)$ for the selected red and the blue galaxies, 
separately. Then, we compare the observational results with the analytic model 
(\ref{eqn:2d}), determining the best-fit values of the two parameters 
$a_{\rm l}$ and $\varepsilon_{\rm nl}$ with the help of the 
$\chi^{2}$-minimization. In the fitting procedure, the values of 
the two parameters are confined to $0\le a_{\rm l}\le 0.6$ and 
$0\le \varepsilon_{\rm nl}\le 0.1$\citep{lee-pen01,lee-pen07}. 
The unmarginalized errors in the fitting parameters are calculated by the 
formula given in \citet{bev-rob96}: The error in the parameter 
$a_{\rm l}$ is found as 
$\sigma_{a}=\Delta a_{\rm l}\sqrt{2(\chi^{2}_{1}-2\chi^{2}_{\rm min} + 
\chi^{2}_{2})}$. Here $\Delta a_{\rm l}$ represents a fitting step-size, 
$\chi^{2}_{\rm min}$ is the minimum value of $\chi^{2}$, 
$\chi^{2}_{1} \equiv \chi^{2}(a_{\rm l0}+\Delta a_{\rm l})$
and $\chi^{2}_{1} \equiv \chi^{2}(a_{\rm l0}-\Delta a_{\rm l})$.
where $a_{\rm l0}$ is defined as 
$\chi^{2}_{\rm min}\equiv \chi^{2}(a_{\rm l0})$. 
The unmarginalized error $\sigma_{\varepsilon}$ in the parameter 
$\varepsilon_{\rm nl}$ is also calculated in a similar manner.

For the analytic model (eq.[\ref{eqn:2d}]), we use the $\Lambda$CDM 
power spectrum given by \citet{bar-etal86}. For the Lagrangian scale $R$ of 
each galaxy, we first calculate the luminosity $L$ of each galaxy from its 
absolute r-band magnitude and then convert it to the galaxy mass scale 
using the relation of $L\sim M^{0.88}$ \citep{val-ost06}. Then, the Lagrangian 
galactic scale $R$ is found as a top-hat radius that encloses the mean mass 
$\bar{M}$ averaged over the galaxies of each sample.
When we calculate the absolute magnitude of each galaxy, no $K$-correction is 
considered: If our goal were to investigate how $\eta_{2D}(r)$ changes in 
narrow absolute magnitude bin, then it would be necessary to determine 
$M_{r}$ accurately by considering the $K$-correction \citep{wak-etal06}. 
However, our goal here is rather to measure the average $\eta_{2D}(r)$ 
for those galaxies which are brighter than a single magnitude threshold 
in the redshift range of $0\le z \le 0.4$. Hence, a maximum $20\%$ error 
that may be caused by ignoring $K$-correction in the estimation of the 
galaxy absolute magnitude should not be an issue here. 

\begin{figure}
  \begin{center}
   \plotone{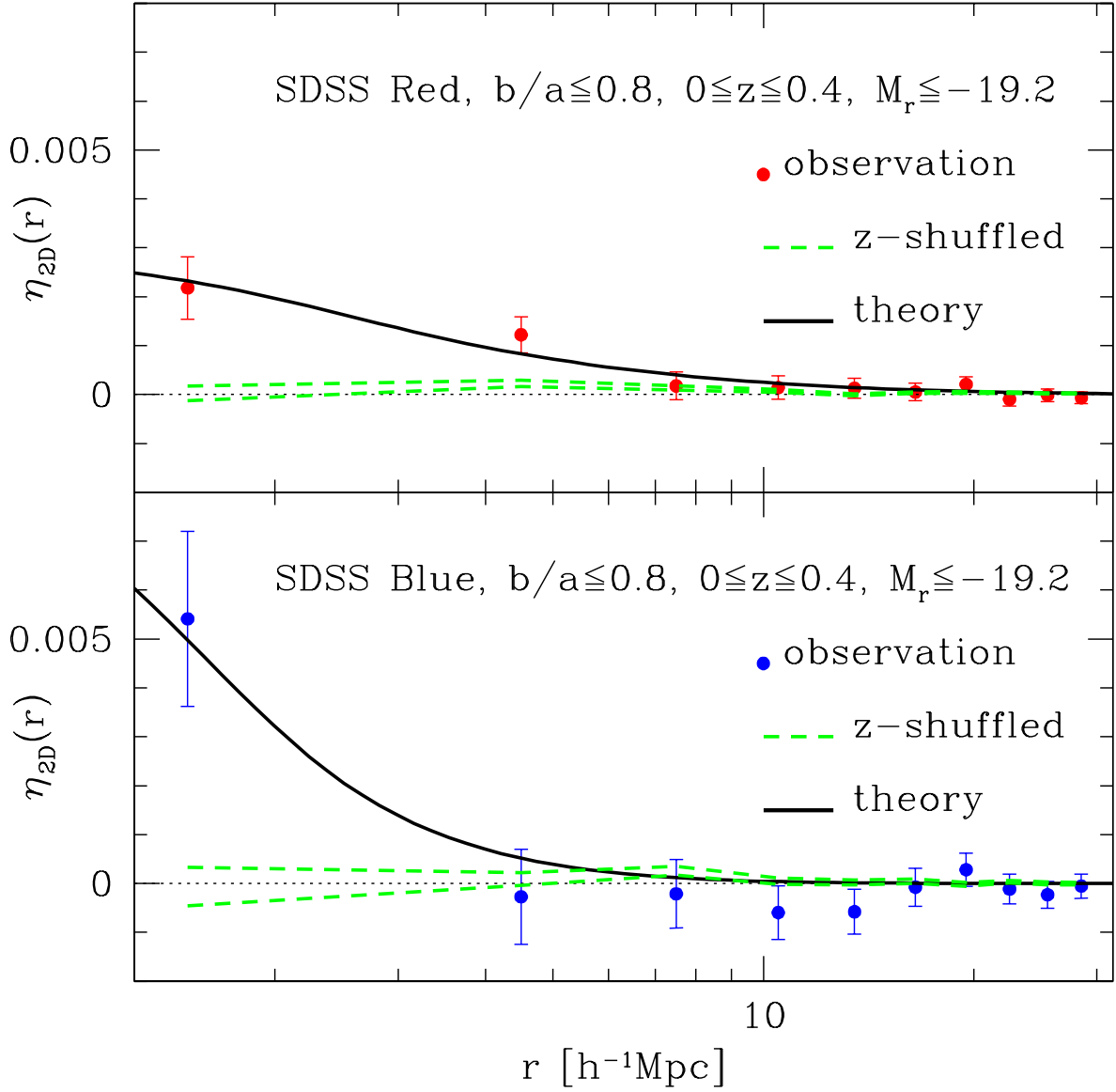}
\caption{Correlations of the two dimensional projected ellipticities from 
the SDSS red and the blue galaxies (top and bottom, respectively) with the 
absolute magnitude cut of $M_{r}\le -19.2$ in the redshift range of 
$0\le z\le 0.4$.  In each panel, the solid dots represent the observed 
signals, while the solid line corresponds to the analytic model with the 
best-fit parameters  (eq.[\ref{eqn:2d}]) found through 
$\chi^{2}$-minimization. The dashed line corresponds to the correlation 
from the data with $z$-shuffled. For the observed signals, only those 
galaxies with major-to-minor axis ratio of $b/a\le 0.8$ are considered.}
\label{fig:cor}
 \end{center}
\end{figure}

Fig. \ref{fig:cor} plots the observational results of $\eta_{2D}(r)$ from the 
red and the blue galaxy sample as solid dots in the top and the bottom panel, 
respectively. In each panel, the solid line represents the analytic model 
with the best-fit correlation parameters $a_{\rm l}$ and $\epsilon_{\rm nl}$.
The errors are calculated as standard deviation for the case of 
no correlation as  $\sigma_{\eta}=1/\sqrt{8n_{\rm p}}$where $n_{\rm p}$ is the 
number of galaxy pairs belonging to each bin \citep{pen-etal00}. As can be 
seen, clear signals of ellipticity correlation higher than $3\sigma_{\eta}$ 
are detected from both the red and the blue galaxy samples at the first 
bin ($r \le 3h^{-1}$Mpc). Note also that from the red galaxy sample the 
correlation signal is also detected at the second distance bin 
($r \le 3h^{-1}$Mpc). 
 
Table \ref{tab:cor} lists the best-fit values of $a_{\rm l}$ and 
$\epsilon_{\rm nl}$. The results show a marked difference between the blue 
and the red galaxy intrinsic alignments: For the blue galaxies, the best-fit 
values of the correlation parameters are found to be $a_{\rm l}=0.20\pm 0.04$ 
and $\epsilon_{\rm nl}\approx0.$, which indicates that $\eta_{\rm 2D}$ of the 
blue galaxies follows a quadratic scaling of $\xi(r)$, as predicted  
by the linear tidal torque theory \citep{pen-etal00,lee-erd07}. 
In contrast, for the red galaxies, the best-fit values are 
$a_{\rm l}\approx 0$ and $\epsilon_{\rm nl}=(2.6\pm 0.5)\times 10^{-3}$, 
which implies that   $\eta_{2D}$ of the bright red galaxies follows a linear 
scaling of $\xi(r)$.  This observational result is consistent with the 
picture that the ellipticities of the blue and the red galaxies are induced 
by the initial tidal field and the anisotropic infall of materials, 
respectively. 

Before assuring ourselves that the detected signal is real, however, it is 
necessary to examine the effect of the systematic error in the measurement 
of the position angles of the SDSS galaxies because the systematic error 
could cause correlations of galaxy ellipticities that mimic intrinsic 
alignments. To quantify the systematic error, we perform two simple tests. 
First, we shuffle randomly the redshifts of the selected SDSS galaxies in 
each sample and remeasure $\eta_{2D}$ one hundred times. 
For each distance bin, we calculate the mean averaged over these $100$ random 
realizations and the standard deviation between realizations, assuming that 
these $100$ shuffling processes are mutually independent. 
If the systematic error had a dominant effect, then a correlation signal would 
not disappear even when the redshifts are shuffled. Fig \ref{fig:cor} plots 
the mean plus and minus one standard deviation as (green) dashed lines. 
As can be seen, when the redshifts of the selected galaxies are shuffled, 
the ellipticity correlations between the galaxies disappear, which suggests 
that the effect of the systematic error on the galaxy intrinsic alignments 
be negligible.

\begin{figure}
  \begin{center}
   \plotone{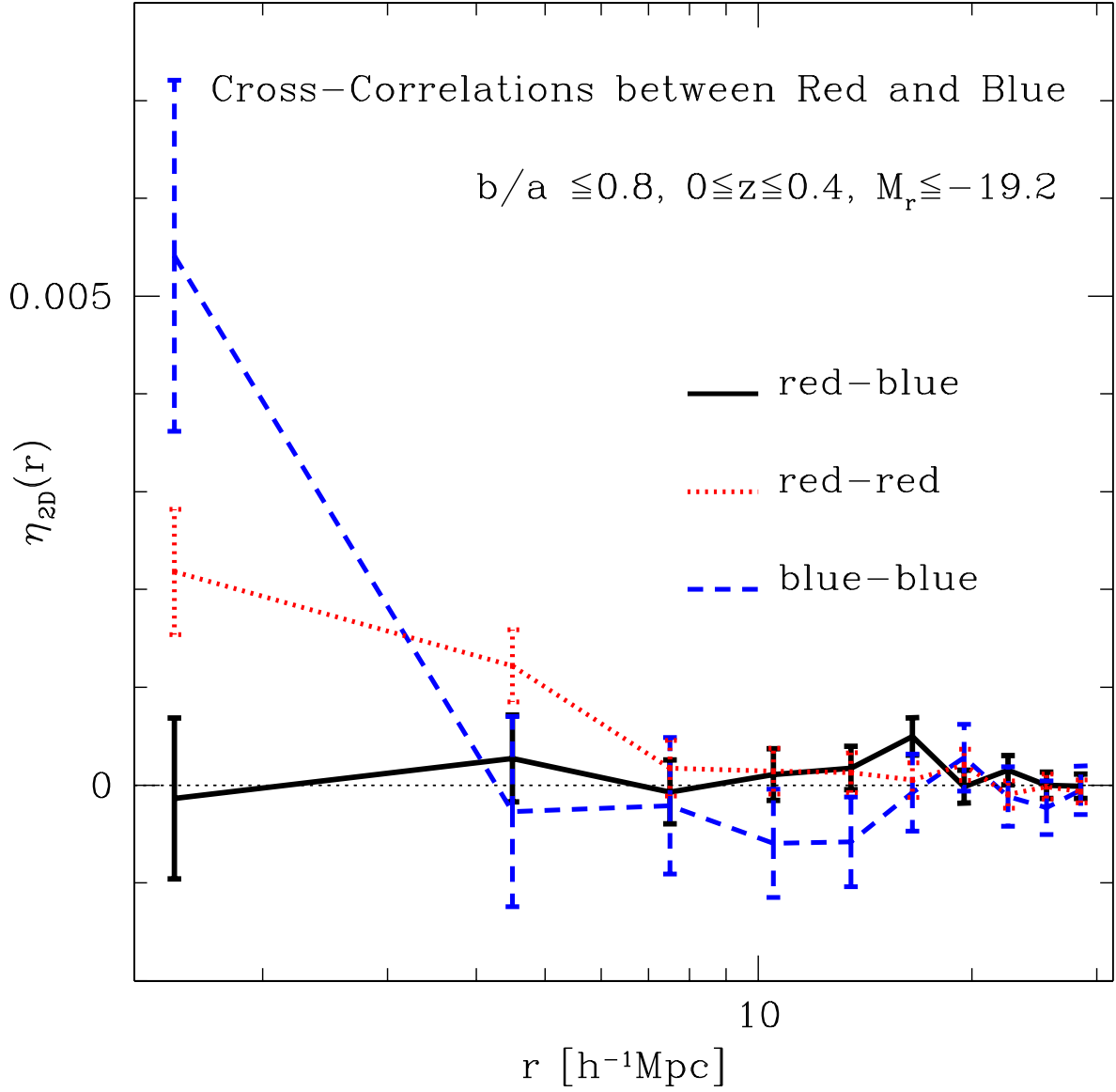}
\caption{Cross-correlations of the two dimensional projected ellipticities 
between the SDSS red and the blue galaxies (solid).}
\label{fig:cross}
 \end{center}
\end{figure}

Second, we measure the ellipticity cross-correlations between the red and 
the blue galaxies, which is plotted as solid line and compared with the 
ellipticity correlation of the red (red dotted) and the blue (blue dashed) 
galaxies in Fig. \ref{fig:cross}. If the systematic error caused the 
correlation of galaxy ellipticities, then we would find non-negligible 
cross-correlations of the ellipticities between the red and the blue galaxies. 
As can be seen in Fig. \ref{fig:cross}, however, the cross-correlations 
are quite weak at the first distance bin, which reassures us that the 
detected ellipticity correlations of the red and the blue galaxies are 
not spurious intrinsic.

It is worth discussing the effect of the redshift distortion that we have 
ignored in our analysis. For the blue galaxies, its mean peculiar velocity 
dispersion has been observed to be quite small less than $150$ km/s on 
average \citep{dav-etal97}. Therefore it is well justified to ignore the 
redshift distortion effect. On the other hand, for the red galaxies 
which are observed to have higher peculiar velocity dispersion around 
$200$ km/s \citep{dav-etal03}, the redshift distortion effect may cause 
non-negligible degree of scatter in the measurement of the galaxy separation 
distance $r$.  Thus, we do not exclude a possibility that the 
characteristic distance scale for the red galaxy intrinsic alignments 
and the value of $\varepsilon_{\rm nl}$ could be smaller in real space 
than observed in redshift space. 

The key implication of our results for the weak lensing survey is also worth 
discussing here. According to our results, the degree of the galaxy intrinsic 
alignments is quite strong for the bright galaxies at redshifts $z\le 0.4$. 
Since the galaxy intrinsic correlations tend to decrease with redshift 
\citep{lee-pen07}, this implies that at $z=1$ relevant for the cosmic shear 
survey the intrinsic correlations may be larger than the detected signal of 
$a_{\rm l}=0.2$ and thus should not be completely negligible. 
Our results also suggest that when removing close galaxy pairs as intrinsic 
ellipticity contaminants from the cosmic shear analysis 
\citep{hey-hea03,tak-whi04}, a typical distance scale of a galaxy 
pair has to be determined with care, since the scaling of the intrinsic 
correlations depend on the intrinsic property of the galaxies (red or blue). 

We conclude that our first detection of a clear signal of the galaxy 
intrinsic correlations from the SDSS data will be useful for the weak 
lensing measurement as well as for understanding the evolution and formation 
of the red and the blue galaxies.

\acknowledgments

We thank an anonymous referee for his/her constructive report which helped 
us improve significantly the original manuscript. We also thank B. M\'enard 
for his help in downloading the SDSS data and R. Mandelbaum for useful 
comments.J.L. is grateful to the warm hospitality of the Canadian Institute 
for Theoretical Astrophysics (CITA) where this project was planned and 
carried out. J.L. acknowledges the financial support from the Korea 
Science and Engineering Foundation (KOSEF) grant funded by the Korean 
Government (MOST, NO. R01-2007-000-10246-0).

Funding for the SDSS and SDSS-II has been provided by the Alfred
P. Sloan Foundation, the Participating Institutions, the National
Science Foundation, the U.S. Department of Energy, the National
Aeronautics and Space Administration, the Japanese Monbukagakusho, the
Max Planck Society, and the Higher Education Funding Council for
England. The SDSS Web Site is http://www.sdss.org/. 

The SDSS is managed by the Astrophysical Research Consortium for the
Participating Institutions. The Participating Institutions are the
American Museum of Natural History, Astrophysical Institute Potsdam,
University of Basel, University of Cambridge, Case Western Reserve
University, University of Chicago, Drexel University, Fermilab, the
Institute for Advanced Study, the Japan Participation Group, Johns
Hopkins University, the Joint Institute for Nuclear Astrophysics, the
Kavli Institute for Particle Astrophysics and Cosmology, the Korean
Scientist Group, the Chinese Academy of Sciences (LAMOST), Los Alamos
National Laboratory, the Max-Planck-Institute for Astronomy (MPIA),
the Max-Planck-Institute for Astrophysics (MPA), New Mexico State
University, Ohio State University, University of Pittsburgh,
University of Portsmouth, Princeton University, the United States
Naval Observatory, and the University of Washington.

\clearpage
\begin{deluxetable}{cccccc}
\tablewidth{0pt}
\setlength{\tabcolsep}{5mm}
\tablehead{
Color & $z$ & $N_{g}$ & $M_{r}$ &  $a_{\rm l}\times 10^{1}$ & 
$\varepsilon_{\rm nl}\times 10^{3}$ \\ 
&&&(no $K$-correlation)&&} 
\tablecaption{The redshift range ($z$), a total number of galaxies 
($N_{g}$), the range of the absolute magnitude ($M_{r}$), the 
best-fit-values of the linear and the nonlinear correlation parameters.}
\startdata
Red & $[0.0,0.4]$ & $283972$ & $\le -19.2$ & $0.0\pm 3.1$ & $2.6\pm 0.5$ \\ 
Blue & $[0.0,0.4]$ & $87188$ & $\le -19.2$ & $2.0\pm 0.4$ & $0.0\pm 1.6$ \\ 
\enddata
\label{tab:cor}
\end{deluxetable}

\clearpage
\clearpage


\begin{thebibliography}{}
\bibitem[Bardeen et al.(1986)]{bar-etal86}
Bardeen, J. M., Bond, J. R., Kaiser, N., \& Szalay, A. S. 1986, \apj, 304, 15
\bibitem[Bevington \& Robinson(1996)]{bev-rob96}
Bevington, P. R., \& Robinson, D. K. 1996, Data Reduction and Error 
Analysis for the Physical Sciences (Boston : McGraw-Hill)
\bibitem[Blanton et al.(2003)]{bla-etal03}
Blanton, M. R., et al. 2003, \apj, 592, 819
\bibitem[Bond, Kofman, \& Pogosyan(1996)]{bon-etal96}
Bond, J., R., Kofman, L., \& Pogosyan, D. 1996, Nature, 380, 603
\bibitem[Brown et al.(2002)]{bro-etal02}
Brown, M. L., Taylor, A. N., Hambly, N. C., \& Dye, S. 2002, \mnras, 333, 501
\bibitem[Catelan et al. (2001)]{cat-etal01}
Catelan, P., Kamionkowski, M., \& Blandford, R. D. 2001, \mnras, 320, L7
\bibitem[Davis et al.(1997)]{dav-etal97}
Davis, M., Miller, A., \& White, S.D.M. 1997, \apj, 490, 63
\bibitem[Davis et al.(2003)]{dav-etal03}
Davis, A. N., Dragan, H., \& Krauss, L. M. 2003, \mnras, 344, 1029
\bibitem[Doroshkevich(1970)]{dor70}
Doroshkevich, A. G. 1970, Astrofizika, 6, 581
\bibitem[Heymans \& Heavens(2003)]{hey-hea03}
Heymans, C. \& Heavens, A. 2003, \mnras, 339, 711
\bibitem[Heymans et al.(2004)]{hey-etal04}
Heymans, C., Brown, M., Heavens, A., Meisenheimer, K., Taylor, A., \& 
Wolf, C. 2004, \mnras, 361, 160
\bibitem[Hirata \& Seljak(2004)]{hir-sel04}
Hirata, C. M. \& Seljak, U. 2004, \prd, 70, 063526
\bibitem[Hirata et al.(2004)]{hir-etal04}
Hirata, C. M., et al. 2004, \mnras, 353, 529
\bibitem[Hirata et al.(2007)]{hir-etal07}
Hirata, C. M., et al. 2007, preprint [astro-ph/0701671] 
\bibitem[Hui \& Zhang(2002)]{hui-zha02}
Hui, L. \& Zhang Z. 2002, preprint [astro-ph/0205512]
\bibitem[Jing(2002)]{jin02}
Jing, Y. 2002, \mnras, 335, 89
\bibitem[Lee \& Pen(2001)]{lee-pen01}
Lee, J. \& Pen, U. L. 2001, \apj, 555, 106
\bibitem[Lee \& Pen(2002)]{lee-pen02}
Lee, J. \& Pen, U. L.  2002, \apj, 567, 111
\bibitem[Lee \& Erdogdu(2007)]{lee-erd07}
Lee, J. \& Erdogdu, P. 2007, preprint [arXiv:0707.1611]
\bibitem[Lee \& Pen(2007)]{lee-pen07}
Lee, J. \& Pen, U. L.  2007, preprint [arXiv:0707.1690]
\bibitem[Lee et al.(2007)]{lee-etal07}
Lee, J. Springel, V., Pen, U. L., Lemson, G. 2007, preprint [arXiv:0709.1106]
\bibitem[Lupton et al.(1999)]{lup-etal99}
Lupton, R. H., Gunn, J. E., \& Szalay, A. S. 1999, \apj, 118, 1406
\bibitem[Mandelbaum et al.(2006)]{man-etal06}
Mandelbaum, R., Hirata, C. M., Ishak, M., Seljak, U., \& Brinkmann, J. 
2006, \mnras, 367, 611
\bibitem[Pen et al.(2000)]{pen-etal00}
Pen, U. L., Lee, J., \& Seljak, U. 2000, 543, L107
\bibitem[Strateva et al.(2001)]{str-etal01}
Strateva, I., et al. 2001, \apj, 122, 1861
\bibitem[Takada \& White(2004)]{tak-whi04}
Takada, M., \& White, S. D. M. 2004, \apj, 601, L1
\bibitem[Vale \& Ostriker(2006)]{val-ost06}
Vale, A., Ostriker, J. P. 2006, \mnras, 371, 1173
\bibitem[York et al.(2000)]{yor-etal00}
York, D. G., et al. 2000, \mnras, 372, 537
\bibitem[Wake et al.(2006)]{wak-etal06}
Wake, D. A., et al. 2006, \mnras, 372, 537
\bibitem[West(1994)]{wes94}
West, M. J. 1994, \mnras, 268, 79
\bibitem[White(1984)]{whi84}
White, S. D. M. 1984, \apj, 286, 38

\end{thebibliography}
\end{document}